\documentclass{emulateapj}

\usepackage{epsfig}

\newcommand{\ltsim}{\protect\raisebox{-0.5ex}{$\:\stackrel{\textstyle <}
        {\sim}\:$}}
\newcommand{\gtsim}{\protect\raisebox{-0.5ex}{$\:\stackrel{\textstyle >}
        {\sim}\:$}}

\newcommand{\msun}{M_{\odot}}
\newcommand{\rsun}{R_{\odot}}
\newcommand{\lsun}{L_{\odot}}
\newcommand{\mc}{m_c}
\newcommand{\sigmac}{\Sigma_c}
\newcommand{\rc}{r_o}
\newcommand{\ms}{m_{*}}
\newcommand{\rs}{r_{*}}
\newcommand{\dms}{\dot{m}_{*}}
\newcommand{\krho}{k_{\rho}}
\newcommand{\kt}{k_T}

\newcommand{\mfr}{m_{\rm frag}}
\newcommand{\rfr}{r_{\rm frag}}

\begin{document}

\title{Radiation Feedback and Fragmentation in Massive Protostellar Cores}

\slugcomment{Accepted for publication in ApJ Letters, \today}

\author{Mark R. Krumholz\footnote{Hubble Fellow}}
\affil{Department of Astrophysics, Princeton University, Princeton, NJ 08544, USA}

\begin{abstract}
Star formation generally proceeds inside-out, with overdense regions inside protostellar cores collapsing rapidly and progressively less dense regions following later. Consequently, a small protostar will form early in the evolution of a core, and collapsing material will fall to the protostellar surface and radiate away its gravitational potential energy. The resulting accretion luminosity will heat the core and may substantially affect the process of fragmentation. This is of particular interest for massive cores that, at their initial temperatures, have masses much greater than a thermal Jeans mass and thus might be expected to fragment into many stars during collapse. Here I show that accretion luminosity can heat the inner parts of a core to $>100$ K very early in the star formation process, and that this in turn strongly suppresses fragmentation. This has implications for a number of outstanding problems in star formation, including the mechanism of massive star formation, the origin of the stellar initial mass function and its relationship to the core mass function, the demographics of massive binaries, and the equation of state in star-forming gas.
\end{abstract}


\section{Introduction}

In the last few years, millimeter and sub-mm observations have reached resolutions sufficient to identify dense molecular condensations with masses $\sim 100$ $\msun$ and sizes $\sim 0.1$ pc in regions of massive star formation \citep[a recent review is given in][]{garay06}. These objects are cold, supersonically turbulent, and centrally condensed. In some cases they show signs of coherent infall \citep{beuther05b}. The condensations have a mass distribution that resembles the stellar initial mass function \citep[IMF,][]{beuther04b,reid05}, which naturally leads to the question of whether they could collapse to form massive stars, making them the high mass analogs of the cores from which low mass stars form.

One objection to this scenario is the possibility of fragmentation. Since the thermal Jeans mass in a cold $(T\sim 10$ K), dense ($n_H\sim 10^6$ cm$^{-3}$) gas is $\sim 1$ $\msun$, why should a core of mass $\sim 100$ $\msun$ collapse to form a single massive object object (or a small-multiple system), rather than many smaller objects? \citet{bate05} argue based on simulations that denser regions produce smaller fragments, since the Jeans mass decreases with density at fixed temperature \citep[although see][]{martel06}. \citet{dobbs05} simulate the collapse of a massive turbulent core and find that it fragments to as many as $20$ and as few as $2$ objects, depending on the assumed equation of state. If fragmentation of massive cores to many objects were common, there would be no direct mapping between the core and stellar mass distributions; their agreement would simply be a coincidence. If, on the other hand, massive cores typically collapse to one or a few objects, then one can explain the IMF in terms of the core mass function. (Fragmentation to $2-3$ objects does not pose a problem in mapping from core to star masses, because the massive star IMF is uncorrected for multiplicity.)

Almost all published work on fragmentation of massive cores treats the gas as either isothermal or barotropic. The barotropic approximation is based on calculations showing that radiative cooling keeps gas isothermal until it reaches a density of $\sim 10^{-14}$ g cm$^{-3}$, at which point it transitions to adiabatic \citep{larson69, masunaga98, masunaga00}. However, this approach assumes that gas is heated solely by compression. While this is probably reasonable far from point sources of radiation, it is likely to fail once a protostar forms, since heating of the gas by irradiation from the luminous, accreting central object vastly exceeds heating due to gas compression. Indeed, in detailed one-dimensional radiation-hydrodynamic calculations of the evolution of a $1$ $\msun$ core, \citet{masunaga00} find that accretion luminosity can heat gas to $>100$ K out to hundreds of AU. \citet{matzner05} find that heating due to accretion luminosity is sufficient to prevent fragmentation of low mass protostellar disks into brown dwarfs.

Massive star-forming regions have high densities that produce accretion rates orders of magnitude larger than in low mass regions, and they are optically thick even in the mid-infrared. Thus, heating by accretion luminosity within them is likely to have even more profound effects. In this paper I test the assumption that massive cores can be described with a barotropic equation of state, and explore how a more realistic treatment of gas thermodynamics changes the picture of fragmentation in these objects. In \S~\ref{tempmodel}, I develop a simple analytic model for the temperature distribution in massive cores, and in \S~\ref{fragmodel} I use the results to estimate how fragmentation is likely to be affected. Finally, in \S~\ref{conclusions} I discuss some of the broader implications of my findings.

\section{The Temperature Structure in Massive Cores}
\label{tempmodel}

Since the focus of this paper is fragmentation, I concentrate on cores
at early times when the gas mass greatly exceeds the stellar mass, and
accretion is the dominant source of luminosity. To make the
calculations definite, I use the turbulent core model proposed by
\citet[hereafter MT03]{mckee03}, in which the
density distribution within a core is a power law, $\rho\propto
r^{-\krho}$, and both MT03 and I adopt $\krho=1.5$ as a fiducial
value. In this model a choice of core mass $\mc$ and mean column density $\sigmac$ specifies the outer radius of the core $r_o=[\mc/(\pi\sigmac)]^{1/2}$ and hence the density profile. Requiring hydrostatic balance gives the effective sound speed versus radius, $\sigma = \sigma_o
(r/r_o)^{1-\krho/2}$, with $\sigma_o = [2(\krho-1)]^{-1/2} (G\mc/\rc)^{1/2}$.
Within the core is an embedded protostar of mass $\ms$. This model neglects
turbulent structure within a core, the effects of which I discuss in more detail in
\S~\ref{fragmodel}.

One can estimate the accretion rate in a massive core from equation (41)
of MT03:
\begin{equation}
\label{mtaccrate}
\dms = 4.6\times 10^{-4} \left(\frac{\mc}{60\,\msun}\right)^{3/4}
\sigmac^{3/4} \left(2\frac{\ms}{\mc}\right)^{1/2} \;\msun\mbox{
yr}^{-1}.
\end{equation}
(This calculation assumes that half the mass reaching the star is
blown away in an outflow.) For a protostar in the mass range $0.01-1$
$\msun$, this gives a typical accretion rate of roughly
$10^{-5}-10^{-4}$ $\msun$ yr$^{-1}$ for a core with $\mc=50$ $\msun$
and $\sigmac=1.0$ g cm$^{-2}$. Though I compute $\dms$
from MT03, one can obtain similar values by simple
order-of-magnitude arguments. For example, \citet{beuther05b} observe
a velocity dispersion $\sigma=1.0$ km s$^{-1}$ in IRDC 18223-3 main, giving an expected
accretion rate $\dms\sim \sigma^3/G \approx 10^{-4}$ $\msun$ yr$^{-1}$. \citet{reid05} find cores with typical values $\mc\sim 100$ $\msun$, $\rc \sim 0.25$ pc. At the time
when the forming protostar has mass $\ms$, the accretion rate should
be $\dms\sim \ms/t_{\rm ff}(\ms)\approx 10^{-5}$ $\msun$ yr$^{-1}$, where $t_{\rm ff}(\ms)$ is the
free-fall time at the edge of the central region in the core that
contains $\ms$ of gas, and the evaluation assumes $\krho=1.5$. Simulations
obtain comparable values. \citet{dobbs05} find accretion rates
onto single objects of $10^{-5}-10^{-4}$ $\msun$ yr$^{-1}$ for
protostars of mass $m_*<\msun$ in a core with
$\mc=30$ $\msun$, $\sigmac=0.6$ g cm$^{-2}$.

To compute the accretion luminosity, one must know the protostellar
radius. Initially 
collapsing gas forms a pressure-supported ``first core" a few AU
in size, but this becomes unstable and collapses to stellar sizes once
its mass reaches roughly $0.05$ $\msun$
\citep{masunaga98,masunaga00}. Thereafter one can compute the
protostellar radius using the simple evolution model
of MT03, which agrees well with the more detailed numerical
calculations of \citet{palla92}. The accretion luminosity is $L_{\rm
acc} = f_{\rm acc} \ms \dms / \rs$, where $f_{\rm acc}\approx 0.5$ is the
fraction of the accretion energy that goes into radiation rather than
driving a wind. For typical
massive star-forming regions, this gives $L_{\rm acc} \sim
10-100$ $\lsun$.

Given a central luminosity and a density structure, one could numerically determine the temperature structure \citep[e.g.][]{ivezic97,whitney03a}. However, for simplicity I instead
use the analytic approximation given by \citet{chakrabarti05}, which
agrees well with numerical calculations. I will not review the full
Chakrabarti \& McKee formalism, but the central idea is that the
majority of the energy radiated away at a given frequency comes from a
thin radial shell whose position is determined by a competition
between opacity, which suppresses emission from small radii, and
decreasing temperature, which reduces emission from large radii. Once
one knows the characteristic emission radius versus frequency, one can
estimate the temperature profile, which is roughly a power law $T=T_{\rm ch} (r/R_{\rm ch})^{-\kt}$. For a given model, I
compute $R_{\rm ch}$, $T_{\rm ch}$, and $\kt$ from Chakrabarti \&
McKee's equations (6), (7), and (41), using the opacity law of
\citet{weingartner01}. Since I am concerned with early times when the
protostar has accreted a negligible fraction of the cloud, I
assume that the core density profile is unchanged from its initial
MT03 form.

\begin{figure}
\epsfig{file=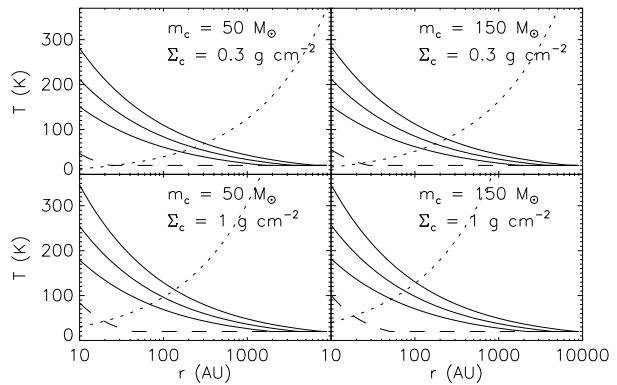}
\caption{
\label{tprof}
Temperature versus radius for models with the values of $\mc$ and $\sigmac$ indicated in the panels. Temperatures shown are for protostellar masses of $\ms=0.8$ $\msun$, $\ms=0.2$ $\msun$, and $\ms=0.05$ $\msun$ (\textit{solid lines, top to bottom}) and for the \citet{dobbs05} barotropic equation of state (\textit{dashed line}). I also show the effective temperature set by the turbulent velocity dispersion, $T_{\rm turb} = 2.33\, m_H \sigma^2 / k_B$ (\textit{dotted line}).}
\end{figure}

I plot the temperature profile for a selection of parameters in Figure \ref{tprof}. The gas is clearly quite warm, but the true temperature is likely even hotter. The \citet{chakrabarti05} approximation underestimates the temperature at very small radii, I have neglected compressional heating, and I have assumed that all of the energy that goes into driving a wind escapes from the core rather than being radiated away by shocks within it. Figure \ref{tprof} also shows the temperature set by the barotropic equation of state used by \citet{dobbs05}, $T=20$ K for $\rho<10^{-14}$ g cm$^{-3}$, $T=431$ K for $\rho>10^{-12}$ g cm$^{-3}$, $T\propto \rho^{2/3}$ in between, which is similar to ones used in other simulations \citep[e.g.][]{bonnell03,bonnell04,bonnell05,bate05}. As
the figure shows, the barotropic
approximation severely underestimates the temperature. One can
understand this intuitively: the barotropic approximation says
that, at some critical density, gas becomes adiabatic and ceases
radiating away the energy it gains through compression. However, since
potential energy varies as $r^{-1}$, the vast majority of the energy
released by a collapse comes out in the final plunge onto the stellar
surface, and is then radiatively transferred to the rest of the
gas. Even when the mass accreted onto a star is a small fraction of
the core mass, the energy released by its fall to very small radii can
dominate the total gravitational potential energy released. By
including only the energy released by the fall from $\rc$ to $\sim 10$
AU, roughly the numerical resolution of most simulations, and not the
energy released from $\sim 10$ AU to $\sim \rsun$, the barotropic
approximation misses the vast majority of the energy of collapse. 

To validate my use of the equilibrium temperature profile, I also
estimate the time required to reach radiative equilibrium. If $E_{\rm
diff}$ is the difference in gas thermal energy between a core whose
temperature is as shown in Figure \ref{tprof} and a core with a
constant temperature 20 K, the time to reach radiative equilibrium is
roughly $t_{\rm rad}=E_{\rm diff}/L_{\rm acc}$. This estimate neglects radiative cooling, but since the cooling rate varies as $T^4$ in optically thick gas, cooling is negligible compared to heating until the gas is very close to equilibrium. For the parameter range $\ms=0.05-1.0$ $\msun$, $\mc=50-150$ $\msun$, and $\sigmac=0.3-1.0$ g cm$^{-2}$, $t_{\rm rad} < 10$ yr. This is shorter than the free-fall time everywhere except in the central few tenths of an AU, so the assumption of radiative equilibrium is justified.

\section{Fragmentation in Massive Cores}
\label{fragmodel}

The high temperatures produced by accretion luminosity will change how
fragmentation proceeds. A cold centrally condensed core will undergo global collapse, but local inhomogeneities produced by turbulence will also be gravitationally amplified, producing a collapsing cluster of small stars, as seen in \citet{dobbs05}, rather than monolithic collapse to a single star. Heating prevents the growth of small perturbations, thereby both reducing the number of perturbations that grow and increasing the mean mass of those that can. To study this effect, I compute the Jeans length $\lambda_J=[\pi c_s^2/(G\rho)]^{1/2}$ as a function of radius $r$
in my model cores. (I take the mean particle mass to be $2.33\, m_H$.)
At radii where $\lambda_J > 2r$, no perturbations that fit within the sphere of radius $r$ are Jeans unstable and no fragmentation can occur. The radius where $\lambda_J=2r$ defines a length
scale $\rfr$ and mass scale $\mfr$ below which there can be no
fragmentation. I compute the Jeans length using the thermal sound speed rather than the effective sound speed set by turbulence because both simulations and theory show that the typical fragment mass in a turbulent medium is the thermal Jeans mass rather than the ``turbulent'' Jeans mass \citep[and references therein]{padoan02,maclow04}. Physically, this occurs because fragment formation in a turbulent medium tends to occur at stagnation points that are relatively non-turbulent.

\begin{figure}
\epsfig{file=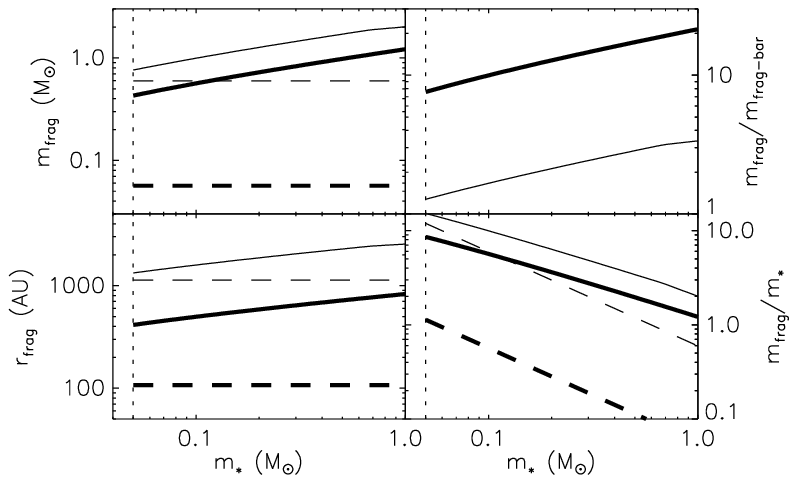}
\caption{
\label{mrfrag}
Mass $\mfr$ and radius $\rfr$ below which fragmentation is suppressed for $(\mc,\sigmac)=(50\,\msun,0.3\mbox{ g cm}^{-2})$ (\textit{thin lines}) and $(150\,\msun,1.0\mbox{ g cm}^{-2})$ (\textit{thick lines}). I plot $\mfr$ (\textit{upper left}), $\rfr$ (\textit{lower left}), the ratio of $\mfr$ computed using radiative models to the value one infers using a barotropic equation of state (\textit{upper right}), and the ratio of $\mfr$ to the protostellar mass $\ms$ (\textit{lower right}). For all plots but the ratio of $\mfr$ computed radiatively and barotropically, I show both radiative transfer (\textit{solid lines}) and barotropic (\textit{dashed lines}) models. The dotted vertical line at $\ms=0.05$ $\msun$ indicates the mass at which a protostar first forms.
}
\end{figure}

Figure \ref{mrfrag} shows $\rfr$ and $\mfr$ for a range of parameter choices. For a barotropic equation of state, there is no heating, so the mass scale is constant. For $\mc=30$ $\msun$, $\Sigma=0.6$ g cm$^{-2}$, the parameters used by \citet{dobbs05}, $\mfr = 0.3$ $\msun$, which roughly agrees with the typical fragment mass in the simulations. However, when one includes radiation, the resulting high temperatures suppress fragmentation to objects of mass $\ltsim 1$ $\msun$ over length scales $\gtsim 1000$ AU from the moment that the first core collapses and a protostar appears. The barotropic approximation underestimates fragment masses by factors as large as $10$, and it switches the problem from the regime where $\mfr > \ms$ (which favors monolithic collapse) to $\mfr<\ms$ (which favors fragmentation).

As illustrated in Figure \ref{tprof},  the velocity dispersion in the radiatively heated gas within a few hundred AU of the star can be larger than the velocity dispersion required to maintain hydrostatic balance. As a result, the gas may temporarily expand, until infall from the colder outer parts of the core raises the density. This could further suppress fragmentation. If the expansion is adiabatic with $\gamma=5/3$, the Jeans length will increase as $\lambda_J\propto \rho^{-1/6}$. With radiative effects, the expansion effectively decreases $\sigmac$ at constant $\mc$, the net effect of which is also to increase $\lambda_J$.

Finally, note that turbulence in massive cores likely produces multiple density peaks within the overall centrally-condensed structure. The densest of these peaks (which is almost certainly close to the center) has the shortest free-fall time and will collapse first. At the accretion rate given by equation (\ref{mtaccrate}) for $\mc=50-150$ $\msun$ and $\sigmac=0.3-1.0$ g cm$^{-2}$, $0.05$ $\msun$ of gas will accumulate, collapse to stellar density, and begin irradiating its surroundings in $30-300$ yr, which is vastly shorter than the mean-density free-fall time of $3\times 10^4-1\times 10^5$ yr. Unless there is an unreasonably synchronized collapse of several density peaks, all the gas in the cloud except the first $0.05$ $\msun$ to collapse will be subject to radiative feedback. In a turbulent core self-shielding may leave sufficiently overdense regions cooler than the spherical calculations indicate, so fragmentation can only be followed in detail by radiation-hydrodynamic simulations. Nonetheless, the spherical calculations definitively show that radiative heating will raise the Jeans length and mass over a large volume within the core well before the vast majority of the gas collapses, so that the number of fragments should be significantly fewer, and their mean mass much larger, than one would estimate by assuming an isothermal or barotropic equation of state.

\section{Implications and Conclusions}
\label{conclusions}

\subsection{Massive Stars, Cores, and the IMF}

The realization that radiative heating can suppress
fragmentation argues that massive condensations really are cores, in
that they are likely to collapse to produce single stars or
small multiple systems, not clusters. This supports the idea that the
massive stars form by accretion (MT03) and that the mass function of
stars is determined at the stage of fragmentation into cores
\citep{padoan02,beuther05b,reid05}. It removes the need to explain the
common shape of the core and stellar mass functions as pure
coincidence.

In contrast, the suppression of fragmentation by radiative heating
presents a serious problem for models of the type proposed by
\citet{bate05} and \citet{bonnell05} in which ejection and competitive
accretion determine the IMF and direct collisions are required to
create massive stars \citep[though see][for a critique of these models
on other grounds]{krumholz05e}. In such models, all stars are born as
brown dwarfs or very low mass stars in clusters $\sim 1000$ AU in
size. Most objects then accrete until being ejected by N-body
interactions, while a few experience runaway collisions, leading to
the formation of massive stars. However, in dense regions where the
accretion luminosity is high, the formation of the first protostar
within a core will inhibit the formation of subsequent ones around
it. Rather than producing a large cluster of low mass protostars or
brown dwarfs as the competitive accretion and collision models demand,
heating will produce a small number of larger protostars.

It is unclear whether these results present a problem to IMF models that depend on the structure of the equation of state for collapsing gas \citep{larson05}. If the typical separation between the fragments in a protocluster is larger than $\sim 1000$ AU, then, as Larson suggests, radiative heating will not substantially modify the typical fragment mass until enough stars have formed to heat the entire protocluster. On the other hand, if the typical inter-fragment spacing is small, then radiative feedback effects cannot be ignored.

\subsection{Binary Formation}

Suppression of fragmentation by feedback also provides a natural explanation for the observation that, when massive stars are in binaries, the binaries tend to consist of two massive stars rather than a high mass star and a low mass one \citep{pinsonneault06}. If binaries result from the fragmentation of massive cores, then one would expect the resulting stars to be massive because, once the first protostar forms, it will raise the Jeans mass and thereby favor the formation of massive stars thereafter. Although I have only considered heating due to accretion luminosity, it is likely to be even more significant once protostars become massive enough to start fusing deuterium and then hydrogen. This will greatly increase their luminosity, and raise the Jeans mass even more. Testing whether this effect can quantitatively explain the observed properties of binaries will require radiation-hydrodynamic simulations. Radiation feedback may also be important for low mass binaries. \citet{lada06} recently pointed out that most low mass stars are singletons rather than binaries. Barotropic simulations of low mass star formation almost invariably produce multiple systems \citep{goodwin04}, and radiation feedback may provide a way of reducing the multiplicity.

\subsection{The Barotropic Approximation in Simulations}

In some cases the barotropic approximation produces results at
odds with radiative transfer calculations even when
there is no point source of radiation \citep{boss00}. Feedback makes the problem far worse. Simulations on scales $\gg 1000$ AU, such as those that model core formation within a larger
molecular cloud or clump \citep[e.g.][]{li04,tilley04}, are probably
safe. On the other hand, as \citet{matzner05} point out for disk
fragmentation, the barotropic approximation is likely to produce
incorrect results in simulations that go to smaller scales
\citep[e.g.][]{bate03, goodwin04}. For this reason, simulations of
fragmentation should be done with radiative transfer. However, in
simulations that do use a barotropic equation of state, one can at
least attempt to check the validity of the approximation ex post facto
by selecting a few time slices and calculating the temperature
distribution, including accretion luminosity, using a radiative
transfer code \citep[e.g.][]{whitney03a}. If the temperature is
significantly higher than indicated by the assumed equation of state
in the simulation, then the results for fragmentation in the region
whose temperature has been underestimated must be regarded with
suspicion.

\acknowledgements I thank I. Bonnell, R. Fisher, R. Klein, C. McKee, J. Stone, and the anonymous referee for helpful comments. This work was supported by NASA through Hubble Fellowship grant \#HSF-HF-01186 awarded by the STScI, which is operated by the AURA, for NASA, under contract NAS 5-26555.





\end{document}